\documentstyle[12pt]{article}
\def\1{\mbox{I\hspace{-.15em}1}}

\def\b{\begin{equation}}
\def\e{\end{equation}}
\def\bee{\begin{enumerate}}
\def\eee{\end{enumerate}}

\title{\emph{Bulk Scale Factor} at Very Early Universe }
\author{M. Mohsenzadeh$^{1}$\thanks{e-mail:
mohsenzadeh@qom-iau.ac.ir} and E. Yusofi$^{2}$\thanks{e-mail:
E.yusofi@iauamol.ac.ir}}
\begin{document}

\maketitle {\it \centerline{\it $^{1}$ Department of Physics, Qom
Branch, Islamic Azad University, Qom, Iran }  \centerline{$^2$
Department of Physics, Amol Branch, Islamic Azad University, Iran}
\centerline{\it P.O.BOX 678, Amol, Mazandaran}}

\begin{abstract}
In this paper we propose a higher dimensional
Cosmology based on FRW model and brane-world scenario. We
consider the warp factor in the brane-world scenario as a scale
factor in 5-dimensional
 generalized FRW metric, which is called as \emph{bulk scale factor}, and obtain the evolution of it with space-like and
time-like extra dimensions. It is then showed that, additional
space-like dimensions can produce exponentially bulk scale factor
under repulsive strong gravitational force in the empty universe
at a very early stage.

\end{abstract}


\section{Introduction and Motivation}
In 1917, a year after Einestein introduced general relativity, he
derived a static cosmological model which was closed \cite{c1}.
His motivation in driving this model was the strong believe that
the universe is static. For this reason, he introduced
cosmological constant in the field equation which is, \b
\Lambda_{0}=8\pi G \rho_{0} \e Where $ G $ is Newton's constant of
gravitation and $ \rho_{0} $ is density of the whole universe.

In the same year, W. de-Sitter obtained another solution to the
modified Einstein's field equation \cite{c2}. The de-Sitter
solution reads as equations \cite{c3} \b
ds^{2}=dt^{2}-a^{2}(t)[dx_{1}^{2}+dx_{2}^{2}+dx_{3}^{2}]=dt^{2}-e^{2Ht}[dx_{1}^{2}+dx_{2}^{2}+dx_{3}^{2}]
\e Where $1$,$2$,$3$ indices refer to the three spatial dimensions
and $ a(t) $ is scala factor and $ H=\sqrt{\frac{\Lambda}{3}} $.
The de-Sitter metric is a solution to homogenous Einstein's
equation, {\it i.e.} for empty universe and
 in the past two decades, observational data have shown that our universe might be in de-Sitter phase.

The most successful model, having wondrous predictive power for
cosmology, was obtained by Friedmann in 1922 \cite{c4}, as well as
by Robertson and Walker in 1935 and 1936,respectively
\cite{c5,c6}, which is called as Friedmann-Robertson-Walker ( FRW
) model. This expanding and centrally symmetric homogeneous model
obey the cosmological principle. The successfulness of this model
is revealed by the fact that the big-bang model is based on this
model \cite{c7,c8}.

After Hubble's observation, in 1929, Einstein realized that he
committed a mistake by introducing the cosmological constant. Due
to this epoch-making observation, cosmologists were prompted to
think for expanding models of the universe more seriously, though
de-Sitter and Friedmann had already derived non-static models by
this time. So, it was natural to think that if the present
universe is so large, it would have been very small and dense in
the extreme past. In 1940, George Gamow addressed to this question
and proposed that, in the beginning of its evolution, universe was
like an extremely hot ball \cite{c9}. He called this ball as
primeval atom and this model is popularly known as the big-bang
model.

On the other hand, the idea that our universe may consist of more
dimensions than the usual 4 dimensional space-time,  was first
considered by T. Caluza in the 1919 \cite{c10,c11,c12}. His aim
was to unify gravity and electromagnetism. Later in 1999, the 5D
(5-dimensional) warped geometry theory, which is a brane-world
theory, developed by Lisa Randall and Ramam Sundrum, while trying
to solve the hierarchy problem of the Standard Model, which is
called RS model  \cite{c13,c14}. They considered one extra
"non-factorizable" dimension and the metric was found for RS model
to be given by \cite{c15,c16,c17}\b
ds^{2}=g_{\mu\nu}a^{2}(y)dx^{\mu}dx^{\nu}-dy^{2}=g_{\mu\nu}
e^{-2ky} dx^{\mu}dx^{\nu}-dy^{2} \e Where Greek indices
$\mu,\nu=(0,1,2,3)$ are refer to the usual four observable
dimensions, and $\textit{y}$ signifies the coordinate on the
additional dimension of length $\textit{r}$ and $ a(y) $ is called
the warp factor and $ k $ is of order of the Plank mass. Here it
is assumed to have two branes. One at $ y=0 $ called Planck-brane
( where gravity is relatively a strong force ) and another at $
y=\pi r $ called the TeV-brane ( where gravity is relatively a
weak force ). In that case, by warping any lagrangian mass
parameter which is naturally $ \approx{M_{PL}} $ ( Planck mass ),
will appear to us in the 4D space-time on the TeV-brane to be $
\approx{TeV} $ (Tera electron-volt). For example, lets see how
this works for the Higgs field on the TeV-brane. First, we right
down the action for the Higgs field in 5D  \b S=\int d^{4}x dy
\sqrt{-g}[
g^{\mu\nu}\partial_{\mu}\hat{H}^{\dag}\partial_{\nu}\hat{H}-\lambda
(\hat{H}^{2}-\nu_{0}^{2})^{2}] \delta(y-\pi r)\e Where $ \nu_{0} $
is the vacuum expectation value of the Higgs field and it is of
order of the Planck mass, and $ \hat{H} $ is Higgs field operator.
The integration over extra dimension is easy to calculate \b
S=\int d^{4}x[ e^{-2kr
\pi}\partial_{\mu}\hat{H}^{\dag}\partial^{\mu}\hat{H}-e^{-4kr \pi}
\lambda (\hat{H}^{2}-\nu_{0}^{2})^{2}]\e If we re-scale the Higgs
field, $ \hat{H}\rightarrow e^{kr \pi}\hat{H} $, this gives:\b
S=\int d^{4}x[
\partial_{\mu}\hat{H}^{\dag}\partial^{\mu}\hat{H}-\lambda (\hat{H}^{2}-\nu_{0}^{2}e^{-2kr\pi})^{2}]\e Thus the Higgs field is
still seen as a Higgs field in four dimensional space-time, but
the vacuum expectation value is now\b
\nu^{2}=\nu_{0}^{2}e^{-2kr\pi} \e Which is at the TeV-scale. Thus
by considering warped extra dimensions one is able to solve the
hierarchy problem. This means that transition from 4D world to 5D
world, exponentially shrink sizes and grows mass and energy
\cite{c18,c19,c20}.

The possibility of the existence of extra dimensions has opened up
new and exciting avenues of research in quantum gravity and
quantum cosmology ( for instance see recently works
\cite{c21,c22,c23,c24,c25} ). Since a higher dimensional world has
more energy ( the range of Planck-scale ) relative to 4D world (
the range of TeV-scale ), in this paper, similar to Gamow idea, we
suggest  higher dimensional world as primeval atom. In contrast to
warped geometry model, which makes transition from 4D world to the
5D world, we go from 5D world to the 4D world. By considering the
symmetry, an expansion in size in our scenario is expected.
Indeed, we intend to propose a higher dimensional Cosmology based
on FRW model and brane-world scenario. For this purpose, We
consider generalized form of FRW metric in 5D space-time and by
solving of the Einstein equation for this 5D space-time, we obtain
the evolution of the bulk scale factor. Thus, the outline of paper
is as follows: In section 2, the scale factor in 4D de-Sitter
space-time is recalled briefly. In section 3, we consider the
evolution of bulk scale factor in 5D space-time with space-like
and time-like extra dimensions. Some conclusions are given in
final section.

\section{Scale factor in 4D de-Sitter space-time}
The standard FRW space-time is\cite{c7} \b
ds^{2}=dt^{2}-a^{2}(t)[dx_{1}^{2}+dx_{2}^{2}+dx_{3}^{2}]. \e Where
$1$,$2$,$3$ indices refer to the three spatial dimensions. In this
case, for empty universe, $ T_{\mu\nu}=0 $, and cosmological
constant, $ \Lambda\neq 0 $, which is the source of gravitation in
place of a massive object, the Enistein's field equation reads  \b
R_{\mu\nu}=\Lambda g_{\mu\nu}. \e  Then we have:\b
-\frac{3\ddot{a}}{a}=\Lambda. \e and \b
a\ddot{a}+2\dot{a}^{2}=\Lambda (-a^{2}). \e Then from (10) and
(11) we have:$$ (\frac{\dot{a}}{a})^{2}=H^{2}=\frac{\Lambda}{3}.
$$ Which integrates to: \b
a(t)=a_{0}e^{Ht}=a_{0}e^{t\sqrt{\frac{\Lambda}{3}}}. \e Where $
a_{0} $ is the scale factor at the end of period inflation in 4D
de-Sitter space-time. So, non-static form of de-Sitter space-time
with exponentially scale factor is obtained as\b
ds^{2}=dt^{2}-a_{0}^{2}e^{2Ht}[dx_{1}^{2}+dx_{2}^{2}+dx_{3}^{2}].\e

\section{Evolution of Bulk scale factor}
By considering the standard FRW metric given by (8) and the metric
(3), generalized form RS brane-world scenario \cite{c15}, we
suggest the generalized form of 5D FRW metric as \b
ds^{2}=g_{ab}dx^{a}dx^{b}=\eta_{\mu\nu}a^{2}(y)dx^{\mu}dx^{\nu}+r^{2}dy^{2},
\e Where \b
g_{ab}=\textit{diag}(a^{2}(y),-a^{2}(y),-a^{2}(y),-a^{2}(y),r^{2}).\e
Which $a,b=(0,1,2,3,4)$ and Greek indices $\mu,\nu=(0,1,2,3)$,
refer to the four observable dimensions, and $\textit{y}$
signifies the coordinate on the additional dimension of length
$\textit{r}$, and scale factor $\tilde{a}=a(y) $, is a
${\textit{bulk scale factor}}$, which only respect to the extra
dimension coordinate $\textit{y}$. For space-like extra dimensions
(\textbf{SLED}), $ r^{2}=-1 $ and for time-like extra dimensions
(\textbf{TLED}), $ r^{2}=+1 $  \cite{c26,c27,c28}.

Similar to section 2, for empty universe which $ T_{ab}=0 $, and
bulk cosmological constant, $ \widetilde{\Lambda}\neq 0 $, the
Enistein's field equation reads as \b R_{ab}=\widetilde{\Lambda}
g_{ab}. \e Then for \textbf{TLED} we have:\b
-3\dot{\tilde{a}}^{2}-\tilde{a}\ddot{\tilde{a}}=\widetilde{\Lambda}
\tilde{a}^{2}, \e Which $ab=(00,11,22,33)$, and for $ ab=44 $, we
have \b \frac{4\ddot{\tilde{a}}}{\tilde{a}}=\widetilde{\Lambda}.\e
Where dot denotes derivative with respect to $y$ as a
\textbf{TLED}.

From Eq.(17) together with Eq.(18), we have:
$$ (\frac{\dot{\tilde{a}}}{\tilde{a}})^{2}=\widetilde{H}^{2}=\frac{-5\widetilde{\Lambda}}{12}
$$ So we obtain, \b \tilde{a}=\tilde{a}_{0}e^{\widetilde{H}y}.\e
Where $ \tilde{H} $ is 5D Hubble parameter, and $ \tilde{a_{0}} $
is the bulk scale factor at the end of its expansion in 5D
space-time. In this case, $ \widetilde{\Lambda} < 0 $, and we have
an expanding space-time with harmonically scale factor (19) as:\b
ds^{2}=\tilde{a}_{0}^{2}e^{2\widetilde{H}y}[dt^{2}-(dx_{1}^{2}+dx_{2}^{2}+dx_{3}^{2})]+dy^{2}.\e

On the other hand, the Enistein's field equation, $
R_{ab}=\widetilde{\Lambda} g_{ab} $, for \textbf{SLED} yields
to:\b 3{\tilde{a}}'^{2}+\tilde{a}\tilde{a}'=\widetilde{\Lambda}
\tilde{a}^{2}.\e where $ ab=(00,11,22,33) $, and for $ ab=44 $ we
have: \b \frac{-4\tilde{a}''}{\tilde{a}}=\widetilde{\Lambda}
(-1).\e Where prime denotes derivative with respect to $ y $ as a
\textbf{SLED}.

Then from (21) and (22), we have:
$$ (\frac{\tilde{a}'}{\tilde{a}})^{2}=\widetilde{H}^{2}=\frac{\widetilde{\Lambda}}{4} $$ Which
integrates to: \b
\tilde{a}=\tilde{a}_{0}e^{\widetilde{H}y}=\tilde{a}_{0}e^{y\sqrt{\frac{\widetilde{\Lambda}}{4}}}\e
Where $ \widetilde{H} $ is 5D Hubble parameter, and $
\tilde{a}_{0} $ is the bulk scale factor at the end of its
expansion in 5D space-time. In this case, $ \widetilde{\Lambda} >
0 $, and we have an expanding space-time with exponentially scale
factor (23) as:\b
ds^{2}=\tilde{a}_{0}^{2}e^{2\widetilde{H}y}[dt^{2}-(dx_{1}^{2}+dx_{2}^{2}+dx_{3}^{2})]-dy^{2}.\e
Though, in the brane picture, the electromagnetism and the weak
and strong nuclear forces are localized on the TeV-brane, but
gravity has no such constraint and so much of its attractive power
"leaks" into the bulk and the force of gravity should appear
significantly stronger on small scales. So, the form of expansion
(24), can be as a consequent of repulsive of this strong
gravitational force in the very early universe with extra
dimensions.

\section{Discussions and Conclusions}
We considered the generalized form of FRW metric similar to
generalized RS brane-world scenario, which admits both
\textbf{SLED} and \textbf{TLED}. We solved the Einstein equation
for an empty 4D and 5D space-time and demonstrated that the
evolution of the bulk scale factor in 5D space-time for
\textbf{SLED} case (24), is exponential form similar to scale
factor in 4D space-time (13). Besides, the following results and
points can be obtained from this work:

$\bullet$ In this scenario, due attention to hierarchy problem and
brane-world Cosmology, it seems more logical that, the 5D very
early universe to have further energy in comparison with the 4D
early universe.

$\bullet$ In the higher dimensional very early universe,
incorporation of hidden extra dimensions and the visible 4D
space-time, the unification of gravitational
force and gauge forces seems to be more possible.

$\bullet$ In this scenario, for TLED we have an harmonically bulk
scale factor (20) and the strong gravitational force is likely
harmonic ( similar to non-static form of anti de-Sitter 5D
space-time ).

$\bullet$ In this scenario, for SLED we have an exponentially bulk
scale factor (24) and the strong gravitational force is likely
repulsive ( similar to non-static form of de-Sitter 5D space-time
). On the other hand, in the brane-world model, by warping any
lagrangian mass, shrink size and the strong gravitational force is
likely attractive. Consequently, in the bulk and in the Planck
scale of energy, gravitation can be attractive as well as
repulsive. Thus, the quantum effects of gravitation are of great
importance \cite{c24}.

$\bullet$ This mechanism of expansion emanate of transition from
unstable and dense state of 5D world ( where is situate on the
exited mode of Kaluza-Klein model and under effects of strong
gravitational force ) to the 4D world ( where is under effects of
weak gravitational force with lesser energy ) \cite{c29}.

These highlight points, will motivate us to suggest another
physics for very early universe in the higher dimensional
Cosmology.

\noindent {\bf{Acknowlegements}}: This work has been supported by the Islamic Azad
University-Qom Branch, Qom, Iran.

\end{document}